
\magnification=\magstep1
\newbox\SlashedBox
\def\slashed#1{\setbox\SlashedBox=\hbox{#1}
\hbox to 0pt{\hbox to 1\wd\SlashedBox{\hfil/\hfil}\hss}#1}
\def\hboxtosizeof#1#2{\setbox\SlashedBox=\hbox{#1}
\hbox to 1\wd\SlashedBox{#2}}

\def\mathslashed#1{\setbox\SlashedBox=\hbox{$#1$}
\hbox to 0pt{\hbox to 1\wd\SlashedBox{\hfil/\hfil}\hss}#1}

\def\ifsmall{\iffalse}  
\def\titlepagefont{}  

\def\DefineTeXgraphics{%
\special{ps::[global] /TeXgraphics { } def}}  

\def\today{\ifcase\month\or January\or February\or March\or April\or May
\or June\or July\or August\or September\or October\or November\or
December\fi\space\number\day, \number\year}
\def\eatPrefix19{}
\def\Year{\expandafter\eatPrefix\the\year}
\newcount\hours \newcount\minutes
\def\monthname{\ifcase\month\or
January\or February\or March\or April\or May\or June\or July\or
August\or September\or October\or November\or December\fi}
\def\shortmonthname{\ifcase\month\or
Jan\or Feb\or Mar\or Apr\or May\or Jun\or Jul\or
Aug\or Sep\or Oct\or Nov\or Dec\fi}

\def\TimeStamp{\hours\the\time\divide\hours by60%
\minutes -\the\time\divide\minutes by60\multiply\minutes by60%
\advance\minutes by\the\time%
${\rm \shortmonthname}\cdot\the\day\cdot\the\year%
\qquad\the\hours:\the\minutes$}




\def\Title#1{%
\vskip 1in{\titlefont\centerline{#1}}\vskip .5in}


\newdimen\fullhsize
\newdimen\hstitle
\hstitle=\hsize 
\newdimen\hsbody
\hsbody=\hsize 
\newdimen\hbodyoffset
\hbodyoffset=\hoffset 
\newbox\leftpage
\def\abstract#1{#1}
\def\rotated{\special{ps: landscape}
\magnification=1000  
\baselineskip=14pt
\global\hstitle=9truein\global\hsbody=4.75truein
\global\vsize=7truein\global\voffset=-.31truein
\global\hoffset=-0.54in\global\hbodyoffset=-.54truein
\global\fullhsize=10truein
\def\DefineTeXgraphics{%
\special{ps::[global]
/TeXgraphics {currentpoint translate 0.7 0.7 scale
              -80 0.72 mul -1000 0.72 mul translate} def}}
\let\lr=L
\def\ifsmall{\iftrue}
\def\titlepagefont{\twelvepoint}
\trueseventeenpoint
\def\almostshipout##1{\if L\lr \count1=1
      \global\setbox\leftpage=##1 \global\let\lr=R
   \else \count1=2
      \shipout\vbox{\hbox to\fullhsize{\box\leftpage\hfil##1}}
      \global\let\lr=L\fi}

\output={\ifnum\count0=1 
 \shipout\vbox{\hbox to \fullhsize{\hfill\pagebody\hfill}}\advancepageno
 \else
 \almostshipout{\leftline{\vbox{\pagebody\makefootline}}}\advancepageno
 \fi}

\def\abstract##1{{\leftskip=1.5in\rightskip=1.5in ##1\par}} }

\global\newcount\secno \global\secno=0
\global\newcount\appno \global\appno=0
\global\newcount\meqno \global\meqno=1
\global\newcount\figno \global\figno=0

\def\Section#1{\global\advance\secno by1\relax\global\meqno=1
\bigbreak\bigskip
\centerline{\twelvepoint \bf %
\the\secno. #1}%
\par\nobreak\medskip\nobreak}
\def\tagsection#1{\edef#1{\the\secno}%
\ifWritingAuxFile\immediate\write\auxfile{\noexpand\xdef\noexpand#1{#1}}\fi%
}
\def\section{\Section}

\def\romappno{\uppercase\expandafter{\romannumeral\appno}}
\def\Appendix#1{\global\advance\appno by1\relax\global\meqno=1\global\secno=0
\bigbreak\bigskip
\centerline{\twelvepoint \bf Appendix %
A. #1}%
\par\nobreak\medskip\nobreak}
\def\tagappendix#1{\edef#1{\romappno}%
\ifWritingAuxFile\immediate\write\auxfile{\noexpand\xdef\noexpand#1{#1}}\fi%
}
\def\appendix{\Appendix}

\def\eqn#1{
\ifnum\secno>0
  \eqno(\the\secno.\the\meqno)\xdef#1{\the\secno.\the\meqno}%
     \global\advance\meqno by1
\else\ifnum\appno>0
  \eqno({\rm\romappno}.\the\meqno)\xdef#1{\romappno.\the\meqno}%
     \global\advance\meqno by1
\else
  \eqno(\the\meqno)\xdef#1{\the\meqno}%
     \glovbal\advance\meqno by1
\fi\fi%
\ifWritingAuxFile\immediate\write\auxfile{\noexpand\xdef\noexpand#1{#1}}\fi%
}
\def\equn{
\ifnum\secno>0
  \eqno(\the\secno.\the\meqno)%
     \global\advance\meqno by1
\else\ifnum\appno>0
  \eqno(A.\the\meqno)%
     \global\advance\meqno by1
\else
  \eqno(\the\meqno)%
     \global\advance\meqno by1
\fi\fi%
}
\def\defeqn#1{
\ifnum\secno>0
  \xdef#1{\the\secno.\the\meqno}%
     \global\advance\meqno by1
\else\ifnum\appno>0
  \xdef#1{\romappno.\the\meqno}%
     \global\advance\meqno by1
\else
  \xdef#1{\the\meqno}%
     \global\advance\meqno by1
\fi\fi%
\ifWritingAuxFile\immediate\write\auxfile{\noexpand\xdef\noexpand#1{#1}}\fi%
}
\def\anoneqn{
\ifnum\secno>0
  \eqno(\the\secno.\the\meqno)%
     \global\advance\meqno by1
\else\ifnum\appno>0
  \eqno({\rm\romappno}.\the\meqno)%
     \global\advance\meqno by1
\else
  \eqno(\the\meqno)%
     \global\advance\meqno by1
\fi\fi%
}
\def\mfig#1#2{\global\advance\figno by1%
\relax#1\the\figno\edef#2{\the\figno}%
\ifWritingAuxFile\immediate\write\auxfile{\noexpand\xdef\noexpand#2{#2}}\fi%
}

\catcode`@=11 

\font\ninerm=cmr9
\font\eightrm=cmr8
\font\sixrm=cmr6

\def\loadtrueseventeenpoint{
 \font\seventeenrm=cmr10 at 17.28truept
 \font\seventeeni=cmmi10 at 17.28truept
 \font\seventeenbf=cmbx10 at 17.28truept
 \font\seventeenit=cmti10 at 17.28truept
 \font\seventeensl=cmsl10 at 17.28truept
 \font\seventeensy=cmsy10 at 17.28truept
}
\def\loadfourteenpoint{
\font\fourteenrm=cmr10 at 14.4pt
\font\fourteeni=cmmi10 at 14.4pt
\font\fourteenit=cmti10 at 14.4pt
\font\fourteensl=cmsl10 at 14.4pt
\font\fourteensy=cmsy10 at 14.4pt
\font\fourteenbf=cmbx10 at 14.4pt
}
\def\loadtruetwelvepoint{
\font\twelverm=cmr10 at 12truept
\font\twelvei=cmmi10 at 12truept
\font\twelveit=cmti10 at 12truept
\font\twelvesl=cmsl10 at 12truept
\font\twelvesy=cmsy10 at 12truept
\font\twelvebf=cmbx10 at 12truept
}

\font\ninei=cmmi9
\font\eighti=cmmi8
\font\sixi=cmmi6
\skewchar\ninei='177 \skewchar\eighti='177 \skewchar\sixi='177

\font\ninesy=cmsy9
\font\eightsy=cmsy8
\font\sixsy=cmsy6
\skewchar\ninesy='60 \skewchar\eightsy='60 \skewchar\sixsy='60

\font\ninebf=cmbx9
\font\eightbf=cmbx8
\font\sixbf=cmbx6

\font\ninett=cmtt9
\font\eighttt=cmtt8

\hyphenchar\tentt=-1 
\hyphenchar\ninett=-1
\hyphenchar\eighttt=-1

\font\ninesl=cmsl9
\font\eightsl=cmsl8

\font\nineit=cmti9
\font\eightit=cmti8


\newskip\ttglue
\def\tenpoint{\def\rm{\fam0\tenrm}%
  \textfont0=\tenrm \scriptfont0=\sevenrm \scriptscriptfont0=\fiverm
  \textfont1=\teni \scriptfont1=\seveni \scriptscriptfont1=\fivei
  \textfont2=\tensy \scriptfont2=\sevensy \scriptscriptfont2=\fivesy
  \textfont3=\tenex \scriptfont3=\tenex \scriptscriptfont3=\tenex
  \def\it{\fam\itfam\tenit}\textfont\itfam=\tenit
  \def\sl{\fam\slfam\tensl}\textfont\slfam=\tensl
  \def\bf{\fam\bffam\tenbf}\textfont\bffam=\tenbf \scriptfont\bffam=\sevenbf
  \scriptscriptfont\bffam=\fivebf
  \normalbaselineskip=12pt
  \let\sc=\eightrm
  \let\big=\tenbig
  \setbox\strutbox=\hbox{\vrule height8.5pt depth3.5pt width\z@}%
  \normalbaselines\rm}

\def\twelvepoint{\def\rm{\fam0\twelverm}%
  \textfont0=\twelverm \scriptfont0=\ninerm \scriptscriptfont0=\sevenrm
  \textfont1=\twelvei \scriptfont1=\ninei \scriptscriptfont1=\seveni
  \textfont2=\twelvesy \scriptfont2=\ninesy \scriptscriptfont2=\sevensy
  \textfont3=\tenex \scriptfont3=\tenex \scriptscriptfont3=\tenex
  \def\it{\fam\itfam\twelveit}\textfont\itfam=\twelveit
  \def\sl{\fam\slfam\twelvesl}\textfont\slfam=\twelvesl
  \def\bf{\fam\bffam\twelvebf}\textfont\bffam=\twelvebf
                                           \scriptfont\bffam=\ninebf
  \scriptscriptfont\bffam=\sevenbf
  \normalbaselineskip=12pt
  \let\sc=\eightrm
  \let\big=\tenbig
  \setbox\strutbox=\hbox{\vrule height8.5pt depth3.5pt width\z@}%
  \normalbaselines\rm}

\def\fourteenpoint{\def\rm{\fam0\fourteenrm}%
  \textfont0=\fourteenrm \scriptfont0=\tenrm \scriptscriptfont0=\sevenrm
  \textfont1=\fourteeni \scriptfont1=\teni \scriptscriptfont1=\seveni
  \textfont2=\fourteensy \scriptfont2=\tensy \scriptscriptfont2=\sevensy
  \textfont3=\tenex \scriptfont3=\tenex \scriptscriptfont3=\tenex
  \def\it{\fam\itfam\fourteenit}\textfont\itfam=\fourteenit
  \def\sl{\fam\slfam\fourteensl}\textfont\slfam=\fourteensl
  \def\bf{\fam\bffam\fourteenbf}\textfont\bffam=\fourteenbf%
  \scriptfont\bffam=\tenbf
  \scriptscriptfont\bffam=\sevenbf
  \normalbaselineskip=17pt
  \let\sc=\elevenrm
  \let\big=\tenbig
  \setbox\strutbox=\hbox{\vrule height8.5pt depth3.5pt width\z@}%
  \normalbaselines\rm}

\def\seventeenpoint{\def\rm{\fam0\seventeenrm}%
  \textfont0=\seventeenrm \scriptfont0=\fourteenrm \scriptscriptfont0=\tenrm
  \textfont1=\seventeeni \scriptfont1=\fourteeni \scriptscriptfont1=\teni
  \textfont2=\seventeensy \scriptfont2=\fourteensy \scriptscriptfont2=\tensy
  \textfont3=\tenex \scriptfont3=\tenex \scriptscriptfont3=\tenex
  \def\it{\fam\itfam\seventeenit}\textfont\itfam=\seventeenit
  \def\sl{\fam\slfam\seventeensl}\textfont\slfam=\seventeensl
  \def\bf{\fam\bffam\seventeenbf}\textfont\bffam=\seventeenbf%
  \scriptfont\bffam=\fourteenbf
  \scriptscriptfont\bffam=\twelvebf
  \normalbaselineskip=21pt
  \let\sc=\fourteenrm
  \let\big=\tenbig
  \setbox\strutbox=\hbox{\vrule height 12pt depth 6pt width\z@}%
  \normalbaselines\rm}

\def\ninepoint{\def\rm{\fam0\ninerm}%
  \textfont0=\ninerm \scriptfont0=\sixrm \scriptscriptfont0=\fiverm
  \textfont1=\ninei \scriptfont1=\sixi \scriptscriptfont1=\fivei
  \textfont2=\ninesy \scriptfont2=\sixsy \scriptscriptfont2=\fivesy
  \textfont3=\tenex \scriptfont3=\tenex \scriptscriptfont3=\tenex
  \def\it{\fam\itfam\nineit}\textfont\itfam=\nineit
  \def\sl{\fam\slfam\ninesl}\textfont\slfam=\ninesl
  \def\bf{\fam\bffam\ninebf}\textfont\bffam=\ninebf \scriptfont\bffam=\sixbf
  \scriptscriptfont\bffam=\fivebf
  \normalbaselineskip=11pt
  \let\sc=\sevenrm
  \let\big=\ninebig
  \setbox\strutbox=\hbox{\vrule height8pt depth3pt width\z@}%
  \normalbaselines\rm}

\def\eightpoint{\def\rm{\fam0\eightrm}%
  \textfont0=\eightrm \scriptfont0=\sixrm \scriptscriptfont0=\fiverm%
  \textfont1=\eighti \scriptfont1=\sixi \scriptscriptfont1=\fivei%
  \textfont2=\eightsy \scriptfont2=\sixsy \scriptscriptfont2=\fivesy%
  \textfont3=\tenex \scriptfont3=\tenex \scriptscriptfont3=\tenex%
  \def\it{\fam\itfam\eightit}\textfont\itfam=\eightit%
  \def\sl{\fam\slfam\eightsl}\textfont\slfam=\eightsl%
  \def\bf{\fam\bffam\eightbf}\textfont\bffam=\eightbf \scriptfont\bffam=\sixbf%
  \scriptscriptfont\bffam=\fivebf%
  \normalbaselineskip=9pt%
  \let\sc=\sixrm%
  \let\big=\eightbig%
  \setbox\strutbox=\hbox{\vrule height7pt depth2pt width\z@}%
  \normalbaselines\rm}

\def\tenbig#1{{\hbox{$\left#1\vbox to8.5pt{}\right.\n@space$}}}
\def\ninebig#1{{\hbox{$\textfont0=\tenrm\textfont2=\tensy
  \left#1\vbox to7.25pt{}\right.\n@space$}}}
\def\eightbig#1{{\hbox{$\textfont0=\ninerm\textfont2=\ninesy
  \left#1\vbox to6.5pt{}\right.\n@space$}}}

\def\footnote#1{\edef\@sf{\spacefactor\the\spacefactor}#1\@sf
      \insert\footins\bgroup\eightpoint
      \interlinepenalty100 \let\par=\endgraf
        \leftskip=\z@skip \rightskip=\z@skip
        \splittopskip=10pt plus 1pt minus 1pt \floatingpenalty=20000
        \smallskip\item{#1}\bgroup\strut\aftergroup\@foot\let\next}
\skip\footins=12pt plus 2pt minus 4pt 
\dimen\footins=30pc 

\newinsert\margin
\dimen\margin=\maxdimen
\def\titlefont{\seventeenpoint}
\loadtruetwelvepoint 
\loadtrueseventeenpoint
\catcode`\@=\active
\catcode`\"=\active

\def\eatOne#1{}
\def\ifundef#1{\expandafter\ifx%
\csname\expandafter\eatOne\string#1\endcsname\relax}


\global\newcount\refno \global\refno=1
\newwrite\rfile
\newlinechar=`\^^J
\def\ref#1#2{\the\refno\nref#1{#2}}
\def\nref#1#2{\xdef#1{\the\refno}%
\ifnum\refno=1\immediate\openout\rfile=refs.tmp\fi%
\immediate\write\rfile{\noexpand\item{[\noexpand#1]\ }#2.}%
\global\advance\refno by1}
\def\lref#1#2{\the\refno\xdef#1{\the\refno}%
\ifnum\refno=1\immediate\openout\rfile=refs.tmp\fi%
\immediate\write\rfile{\noexpand\item{[\noexpand#1]\ }#2\semi}%
\global\advance\refno by1}
\def\cref#1{\immediate\write\rfile{#1\semi}}

\def\semi{;\hfil\noexpand\break}

\newif\ifWritingAuxFile
\newwrite\auxfile
\def\SetUpAuxFile{%
\xdef\auxfileName{\jobname.aux}%
\openin1 \auxfileName \ifeof1\message{No file \auxfileName; I'll create one.
}\else\closein1\relax\input\auxfileName\fi%
\WritingAuxFiletrue%
\immediate\openout\auxfile=\auxfileName}

\def\L{\left(}

\def\tr{\hbox{\rm tr}}
\def\mod{\hbox{\rm mod}}

\def\La{\Lambda}
\def\la{\lambda}

\def\quotesl{\lq\lq}
\def\quotesr{''}

\def\ref#1#2{\nref#1{#2}}
\overfullrule 0pt
\hfuzz 52pt
\hsize 6.50 truein
\vsize 9.5 truein

\loadfourteenpoint
\newcount\meqno
\newcount\secno
\meqno=0
\secno=0
\def\secta{\global\advance\secno by1
\meqno=1}

\def\put#1{\global\edef#1{(\the\secno.\the\meqno)}     }

\def\cgko{1}
\def\cbpz{2}
\def\cfms{3}
\def\ccardya{4}
\def\cgepwitt{5}
\def\ckac{6}
\def\cgod{7}
\def\cgrovel{8}
\def\cmoore{9}
\def\cshellb{10}
\def\czam{11}
\def\cverlinde{12}
\def\cshella{13}
\def\cintr{14}
\def\cbais{15}
\def\cdate{16}
\def\ckaca{17}
\def\ccappelli{18}
\def\cravanini{19}
\def\cextended{20}
\def\cfuture{21}

\def\ibar{{\bar\imath}}
\def\bari{\ibar}
\def\jbar{{\bar\jmath}}

\def\L{\Lambda}

\def\asu{\widehat{su}}
\def\calm{{\cal M}}
\noindent   \hfill LTH-92-292 \break

\noindent   \hfill SWAT-92-02 \break

\noindent \hfill  \today \break

\vskip - 1 cm

\Title{\vbox{\centerline{ Characters For}
\centerline{ Coset Conformal Field Theories }}}

\vskip .5 cm
\centerline{\bf David C. Dunbar }

\centerline{\it Dept. of Physics}
\centerline{\it University College of Swansea}
\centerline{\it Swansea, Wales, U.K.}

\vskip 0.5 truecm

\centerline{\bf and }

\vskip 0.5 truecm

\centerline{ \bf  Keith G. Joshi }

\centerline{\it D.A.M.T.P.}
\centerline{\it University of Liverpool }
\centerline{\it Liverpool, L69 3BX, U.K.}

\vskip 1.0 cm

\vskip 2.0 truecm \baselineskip12pt

\centerline{\bf Abstract }

\vskip 0.3 truecm
{  We present formulae for the characters of coset conformal
field theories and apply these to specific examples
to determine the integer shift of
the conformal weights of primary fields. We also present an
example of a coset conformal field theory which cannot
be described by the identification current method.}

\narrower\smallskip

\baselineskip14pt

\vfill

\break



\vskip .4 truecm
\secta
\noindent
{\bf \the\secno . Introduction}

The coset construction [\cgko] of conformal field theories (CFTs)
[\cbpz,\cfms] has
proved to be a practical method for constructing
rational conformal field theories. Indeed it may be that all rational
CFTs have a coset realisation.  Thus the properties
of coset conformal
field theories are important in understanding the general
structure of rational CFTs.  In this paper we investigate various aspects of
coset models. In particular,
the spectrum of a conformal field theory is determined by the
characters [\ccardya,\cgepwitt], and much information can be gleaned by their
examination.
We provide character formulae for a wide class of coset theories
and examine these formulae in specific examples.
We also use the character formulae to deduce
formulae for the ``integer shifts'' of the conformal weights
of primary fields.

In coset CFTs $\hat g/\hat h$,
where $\hat g$ and $\hat h$ are affine algebras [\ckac,\cgod],
the primary fields are labeled by
$\phi^{\La}_{\la}$,
where $\La$ and $\la$ are weights of the Lie algebras $g$ and $h$ respectively.
The conformal weight of $\phi_\la^\La$ is
given by $h=h_\La-h_\la +n$ where $n$ is an integer
known as the integer shift. The evaluation of the
integer shifts is crucial in determining the spectrum of CFTs, which
 is important,
for example, in obtaining the physical spectrum in four
dimensional superstring models [\cgrovel].
Not all pairs of labels $\La$ and $\la$
give genuine and distinct primary fields but some combinations do
not correspond to fields present in the coset, and some
combinations of labels are equivalent [\cmoore].
In examining the spectrum of primary fields in a coset CFT, the
procedure, due to Schellekens and Yankielowicz [\cshellb], for
specifying which fields are non-zero and distinct, by the introduction
of an ``identification'' current, has proved extremely useful.
However, by examining the specific example
$\widehat{su}(3)_2/\widehat{su}(2)_8$
we show that this procedure is not always applicable, contrary to various
suppositions, and for this
\lq\lq maverick'' coset the spectrum of valid inequivalent
fields is not determined solely by an identification current.

The plan of this paper is as follows: In section~2 we review
the coset formulation of Goddard, Kent and Olive of rational
conformal fields theories [\cgko]
and also the Schellekens and Yankielowicz procedure [\cshellb]
for determining the spectrum of  primary fields in such models.
In section~3 and the appendix we derive explicit
character formulae for these theories.
In section~4 we use these formulae to obtain
expessions for the integer shifts in
the case
$\asu(2)_{k_1}\times \asu(2)_{k_2}/ \asu(2)_{k_1+k_2}$.
In sections 5 and 6 we use these formulae to
obtain information on the conformal field theory for
specific examples namely, $(\hat g_2)_k/\asu(3)_k$ and
$\asu(3)_k/\asu(2)_{4k}$.
The case $\hat g_2/\asu(3)$ gives an example of how
the coset CFT may have a larger symmetry than the Virasoro algebra. In
this
case the coset provides a realisation of the $W_3$ algebra [\czam].
The case $\asu(3)_2/\asu(2)_8$
is very interesting because is provides an example of
a coset CFT which {\it cannot} be described in terms of an identification
current. This provides a counter example to several propositions regarding
coset CFTs.



\vskip .4 truecm
\secta
\noindent
{\bf \the\secno . Review of the GKO construction}

Here we briefly review the Goddard, Kent and Olive (GKO) [\cgko]
construction for rational conformal field theories
and the Schellekens-Yankielowicz [\cshellb] method for
describing primary fields of coset models by the use of
an identification current.

Consider a Kac-Moody algebra $\hat g$ [\ckac],
associated with the Lie algebra $g$.
The corresponding current algebra has fields $J^a(z)$
which satisfy the operator
product expansion (O.P.E.)
$$
J^a (z) J^b(w)
= { -k \delta^{ab} \over (z-w)^2 }
+{ f^{ab}{}_c J^c(w) \over (z-w) }   +{\rm n.s.t.}
\equn
$$
where the integer $k$ is the level of the algebra, $f^{ab}{}_c$ are
the structure constants of the Lie algebra and \lq${\rm n.s.t.}$' denotes
terms which are non-singular as $z\rightarrow w$.  The Laurent coefficients
of the currents $J^a(z)=\sum_{n} J^{a}_n z^{-n-1}$, satisfy
$$
[ J^a_m , J^b_n ] =
f^{ab}{}_c J^c_{m+n}  +  k m \delta^{ab}\delta_{m+n,0}
\; .
\equn
$$
As is well known,
Kac-Moody algebras contain Virasoro algebras.
The stress-energy tensor is formed using the
Sugawara construction [\cgod]
$$
T_g(z) =
{-1 \over  2( k+g) } \sum_a  : J^a J^a (z) :
\equn
$$
which satisfies
$$
T_g (z)T_g(w)= { c_g /2 \over (z-w)^4 } +
{ T_g (w) \over (z-w)^2 }
+ { \partial T_g (w) \over (z-w) }
+ {\rm n.s.t.}
\equn
$$
as $z\rightarrow w$. The central charge is related to
the level of the Kac-Moody algebra by
$$
c_g = { k \dim g  \over k+g }
\equn
$$
where $g$ is the dual Coxeter number of the Lie algebra $g$.

Suppose $g$ contains a subalgebra $h$. Then $\hat g$ will contain
a subalgebra $\hat h$.
We can then form the stress-energy tensor  [\cgko]
$$
T_{g/h} = T_g - T_h
\equn
$$
which satisfies the O.P.E. for an energy momentum
tensor with central charge
$$
c_{g/h} = c_{g}- c_{h}
\equn
$$
This construction of the energy momentum tensor is the
well known GKO construction for coset conformal field theories.
An important feature for this construction is that the O.P.E. of the
currents of the subalgebra $J^{\bar a} (z)$
with $T_{g/h} (w)$ is non-singular
$$
J^{\bar a} (z) T_{g/h} (w) = {\rm n.s.t.}
\equn
$$
or, equivalently,
$$
[ J^{\bar a }_m , L^{g/h}_n ] = 0
\equn
$$
Using this construction it is possible to construct
many CFTs with relatively small central charge, and it may be
that all rational CFTs have a coset realisation.
Representations of $\hat g$ at level $k$ are labeled by the highest
weights of $g$, $\Lambda$, satisfying $\psi\cdot \Lambda \leq k$ where
$\psi$ is the highest root of $g$. The conformal weight of the
associated primary field is given by
$$
h_{\Lambda} =
{ \Lambda^2 + 2\La\cdot\rho_{g} \over
2( k+g)  }
\equn
$$
where $\rho_g$ is half the sum of the positive roots of $g$.

If we consider $\hat g$ as the algebra generated by $J^{a}_n$, then this
contains, by definition, the algebra, $\hat h$ and the algebra generated
by $L^{g/h}_n$. However the full algebra of the coset will be the
algebra $A_{g/h}$, which we define to be the largest algebra
such that
$$
  A_{g/h} \otimes \hat{h} \subset \hat{g}
\equn
$$
In general, a representation of $\hat g$ will decompose into
a sum of representations of the smaller algebra $A_{g/h}\otimes\hat{h}$,
$$
\Lambda =
\sum_{\lambda}  R^{\Lambda}_{\lambda} \otimes \lambda
\equn
$$
where $R^{\La}_{\la}$ is a representation of $A_{g/h}$.

For a representation $\Lambda$ of the Kac-Moody algebra the
character is defined as
$$
\eqalign{
ch_{\Lambda}(\tau,z^i)
&\equiv\tr(exp(\{2\pi
  i\tau L_0+{c\over24}-h_\La +\sum_i z_i H^i_0  \} ) )\cr
&= \tr ( q^{L_0+c/24-h_\La}  \prod_i w_i^{H_0^i} )\cr}
\equn
$$
where $H^i$ are the elements of the Cartan subalgebra of $g$,
$q\equiv exp(2\pi i\tau)$ and $w_i\equiv exp(2 \pi i z_i)$.
The restricted characters $\chi_\La(\tau)$ are defined by
$$
\chi_\La(\tau)\equiv q^{h_\La-c/24}ch_\La(\tau,0)
\equn
$$
Assuming, for the moment,
that the Cartan subalgebra has the same dimension
for both $g$ and $h$, i.e. they are of the same rank, then we will have
$z_i H_0^i = z_{\bar i} H_0^{\bar i}$
and, since $L_0^g= L_0^{g/h}+L_0^h$
$$
ch_{\L}(\tau,z_i)=\sum_\lambda b_\lambda^\L(\tau)ch_\lambda(\tau,\bar z_i)
\put\EQSWone\equn
$$
and for the restricted characters
$$
\chi_{\Lambda} ( \tau )
=\sum_{\lambda}  \chi^{\Lambda}_{\lambda} ( \tau )
\; \chi_{\lambda} ( \tau  )
\equn
$$
{}From this we see that
$$
h_\lambda^\L=h_\L-h_\lambda+n
\equn
$$
where $n$ is the order of the leading term in the power
series expansion of $b^\L_\lambda(q)$.
There are many cases where the rank of $g$ and $h$ are different
(in particular the important example $\hat g\times \hat g / \hat g$),
 but these formulae
 still stand, provided we choose the $z_i(z_{\bar i})$ such that
$z_i H^i_0$ can be written as a linear combination of the
$ z_{\bar i}H^{\bar i}_0$, which is equivalent to
specifying the embedding.
The functions $b^{\Lambda}_{\lambda}(\tau)$ are the
\lq\lq branching functions\quotesr\
of the coset CFT.
Not all combinations
of labels $\La$ and $\la$ give rise to non-zero branching functions and not
all distinct labels give rise to distinct branching functions [\cmoore].
Which possibilities are non-zero and are distinct has been resolved
by the introduction [\cshellb] of the concept of an
\quotesl identification current\quotesr\
in the coset  CFT.

The Schellekens-Yankielowicz mechanism [\cshellb]
for deciding which
fields are non-zero and inequivalent is to use an \lq\lq identification
current\quotesr\
which is defined in terms of simple currents of the factors $\hat g$
and $\hat h$. Before discussing the identification current we
define simple currents and the relation to non-diagonal modular
invariants.
A simple current of a general CFT, $J$,
is a primary field with the simple fusion rules
[\cbpz,\cverlinde]
$$
 J \cdot \phi = \phi'
\equn
$$
That is,  for a primary field the
operator product of $J$ with the field only
yields fields from a single conformal family.
If we consider the fusion of $J$ with itself we obtain
a field $J^2$ etc. For a rational CFT with a finite
number of primary fields there must be an integer
$N$ such that $J^N=1$.
In general the action of $J$ upon a primary field
$\phi$ will yield  fields $\{J^r \phi, r=0,1,\dots, N_\phi-1\}$,
where $J^{N_\phi}\phi=\phi$. The
integer $N_\phi$ must be a divisor of $N$.
If a field satisfies $J\cdot\phi=\phi$ then this field is said to be
a fixed point w.r.t $J$.  The existence of fixed points
(and fields with $N_\phi<N$) does a great deal to complicate the
analysis.
For the most part the difficulties associated with fixed points
have been resolved in ref.[\cshellb ] and we will not linger
on their resolution.  The case of most interest to us
 will be when $J$ has integer conformal weight, that is $h(J)$ is integer.
(In general it can be shown that $h(J)=r/N ,\mod(1)$). In this case,
the CFT has a non-diagonal modular invariant (NDMI) [\cshella,\cintr]
\def\vM{{\vphantom{M}}}
$$
\eqalign{
Z =&
\sum_{\phi:h(J\cdot\phi)-h(\phi) \in Z}
{N \over N_{\phi} }
|  \chi^{\vphantom{M^2}}_{\phi}
+ \chi^{\vphantom{M^2}}_{J\cdot\phi} + \cdots
+ \chi^{\vphantom{M^2}}_{J^{N_{\!\phi}-1}\cdot \phi}  |^2
\cr
=&
\sum_{\phi:h(J\cdot\phi)-h(\phi) \in Z}
{N \over N_{\phi} }
\biggl| \sum_{r=0}^{N_{\phi}-1}
\chi^{\vphantom{M^2}}_{J^r_{\vM}\cdot \phi} \biggr|^2
\* .
\cr}
\put{\EQform}\equn
$$
This form has been suggested to be the diagonal modular invariant
of an extended
algebra [\cmoore,\cshella,\cintr,\cbais]
whose characters are $\chi_{\phi}^{\vM}+\chi_{J\cdot\phi}^{\vM}+\cdots$.
Most Kac-Moody algebras contain simple currents,
some examples of which are
given in following sections.
For cases where $N_{\phi} < N$ the extended algebra appears to have
a multiplicity of primary fields with identical characters.

The relationship between the characters of the coset algebra
and the branching functions has been the source of some confusion,
but has been elegantly resolved by Schellekens and Yankielowcz [\cshellb].
This relies on the observation that the diagonal combination of
characters,
$$
Z= \sum_{a} \chi_a \chi_a^*
\equn
$$
where the summation runs over all genuine chararacters of the coset CFT,
must be modular invariant. Since the $\chi_a$ can be expressed as a sum
of the branching functions this can be rewritten as
$$
Z= \sum_{a} | \sum_{\La,\la} n_{\La,\la}^a b^{\La}_{\la} |^2
\equn
$$
Hence one should look for modular invariant combinations of the
branching functions of this form
(which can be recognised as that generated
by a simple current.)
To look for such modular invariants, note that the
branching functions of $\hat g/\hat h$ transform
as the characters of $\hat g\times \hat h^*$.
Hence if one can find a suitable modular invariant for
$\hat g \times \hat h^*$ then the corresponding object
for $\hat g / \hat h$ will be modular invariant and a candidate for
the diagonal modular invariant of the coset.

The critical observation
is that it is possible to find such a current which
generates a modular invariant corresponding to the characters
of $\hat g/\hat h$.
This current, denoted $J_I$, is called the identification current.
After determining
$J_I=\phi^{J^1}_{J^2}$, the non-zero branching functions are those
which have
$$
h( J_I \cdot \phi^{\La}_{\la}  )
- h( \phi^{\La}_{\la}  ) = 0 \hskip 0.5 truecm \mod(1)
\equn
$$
and we have the following equivalence
$$
\phi^\La_\la \equiv
\phi^{J^1\cdot \La}_{J^2\cdot \la}
\equn
$$
The details of the identification current,
for a variety of cosets is given in [\cshellb].
As a simple example, for $\asu(2)_k$ the simple current is $(k)$ which
satisfies $(k)\cdot(l)= (k-l)$. For $k$ odd there are no fixed points.
For $\asu(2)_k\times \asu(2)_1/ \asu(2)_{k+1}$ (a realisation of the
minimal models) the fields are $\phi^{l_1,l_2}_{l_3}$. The condition
for the branching function to be non-zero reduces to
$l_1+l_2-l_3 =0 \quad\mod(2)$, and we have the equivalence
$$
\phi^{l_1,l_2}_{l_3} \equiv \phi^{k-l_1,1-l_2}_{k+1-l_3}
\equn
$$
which we see can be
rearranged as the standard labelling for the fields of the
minimal models.

The identification current method is elegant and applies to many cosets,
however, as we shall show in this paper, there do exist cosets which can
not be described in terms of such an identification current.


\vskip .4 truecm
\secta
\noindent
{\bf \the\secno . Character Formulae }

In this section we explore and determine the branching
functions for coset CFTs. The branching functions are determined
by equ.~\EQSWone .
In general, one cannot set $z^i=0$ to solve these equations
because there is then not enough information due to the
multiplicity of terms on the R.H.S.    The solution to the
special case $\asu(2)_{k_1}\times\asu(2)_{k_2}/\asu(2)_{k_1+k_2}$
has been presented in ref.[\cdate].
Throughout this section we follow the notation of ref [\cgepwitt].
The Weyl-Kac formula for the characters of a Kac-Moody algebra is
$$
ch_{\La,k}^g(\tau,z_i) ={N_{\La,k}^g(\tau,z_i) \over D_g (\tau,z_i)}
\put\EQSWtwo\equn
$$
where
$$
 N^{g}_{\La,k}(\tau,z_i) =
\sum_{w\in W_g}\epsilon(w) \Theta_{w(\La+\rho),k+g } (\tau,z_i)
\put\EQBAKa\equn
$$
and
$$
D_g(\tau,z_i) =
\prod_{{\alpha > 0} \atop {\alpha \in \hat g}} ( 1-e^{-\alpha} )
^{\rm mult (\alpha)}(\tau,z_i)
\put\EQdefch\equn
$$
The $w$ are elements of the Weyl group of the Lie algebra, $W_g$,
and $\epsilon(w)=\pm 1$ is the parity of $w$.
The $\Theta_{w(\La+g),k+g}$
are the generalised $\Theta$-functions
$$
\Theta_{\La,k}( \tau,z_i) =
e^{2\pi i (  \rho \cdot z_i e_i )} \sum_{ \gamma \in M}
q^{ {k \over 2} \gamma\cdot\gamma  + \gamma\cdot\La  }
\exp\Bigl\{ -2\pi i \bigl(k\gamma\cdot (z_i e_i)+\La\cdot (z_i e_i)\bigr)
\Bigr\}
\equn
$$
where $\{e_i\}$ form a basis for the weight lattice of $g$ and
$M$ is the long root lattice of $g$ (which is the root lattice whenever
$g$ is simply laced.)

The denominator in eqn.\EQdefch\
is defined in terms of $\alpha$ where $\alpha$ are all the positive
roots of the Kac-Moody algebra.
The denominator can be rewritten (or is defined as) as
$$
\eqalign{
D_g(\tau,z_i)
&= \cr
\Bigl( &\prod_{n=1}^{\infty} ( 1- q^n) \Bigr)^{{\rm rank}(g)}
\prod_{\bar\alpha \in g^+ } ( 1- e^{2\pi i \bar\alpha\cdot z_ie^i} )
\Bigl( \prod_{n=1}^{\infty}
(1- q^n
e^{2\pi i \bar\alpha\cdot z_ie^i} )
(1-q^n
e^{- 2\pi i \bar\alpha\cdot z_ie^i} )
\Bigr) \cr}
\equn
$$
where $\bar \alpha$ are the roots of the Lie algebra,
and $g^+$ denotes the set of postive roots of $g$.

In this section, for clarity, we restrict ourselves to an illustrative case,
leaving the rather complex general case to the appendix.
Suppose we have a pair of Lie algebras $h$ and $g$, of the same rank
such that the root
lattice of $h$ is contained in that of $g$.
Writing eq.~\EQSWone\ using eq.~\EQSWtwo ,
$$
{N_\La^g(q,z)\over D_g}=
\sum_\mu b_\La^\mu(q){N_\mu^h(q,z)\over D_h}
\; .
\equn
$$
The denominators are defined in terms of a product over the
roots.
For the simple case
examined here
( since the roots of $h$ are a subset of the roots of $g$),
it is easy to construct $D_{g/h}\equiv D_gD_h^{-1}$:
$$
D_{g/h}=\prod_{\alpha\in S}(1-e^{-\alpha})^{{\rm mult }\alpha}
\equn
$$
where $S$ consists of positive roots of $\hat g$ which are not roots of
$\hat h$.
Then
$$
{N_\La^g(q,z)\over D_{g/h}(q,z)} =
\sum_\mu b_\La^\mu(q){N_\mu^h(q,z)}
\put\EQSWthree\equn
$$
Consider the coefficient, on the r.h.s., of
$$
exp(-2\pi i\la\cdot z)
\equn
$$
where $\la$ labels some highest weight irreducible representation
of $\hat h$.
{}From (3.2) and (3.4), there is a unique term of this form,
i.e. that which has $\gamma=0,\omega=1,\mu=\la$, and moreover its
coefficient is the branching function. Thus by comparing coefficients
on both sides of eq.~\EQSWthree\
$$
b_\La^\la=exp(2\pi i\la\cdot z) { N_\La^g(q,z)\over D_{g/h}(q,z)}
\biggr|_{\rm coeff\ of\ \zeta_i^0}
\equn
$$
where $\zeta_i\equiv exp(2\pi iz_i),i=1\dots {\rm rank}\ g$.
The branching
function can then be
obtained by expanding the r.h.s. as a Laurent series in $\{\zeta_i\}$.
The numerator $N_\La^g$ is already in
this form and we can expand the denominator $D_{g/h}$ using
the identity [\ckaca]:
$$
\eqalign{
{ 1 \over
(1-t) \prod_{n=1}^{\infty}  ( 1 -q^n t) (1-q^n t^{-1} )
} =&
\phi(q)^2
\sum_{n=-\infty}^{\infty} (-1)^n
{ q^{ n(n+1)/2}  \over ( 1-q^n t )
}\cr
=&
-\phi(q)^2
\sum_{n=-\infty}^{-1} (-1)^n
 q^{ n(n+1)/2}
\sum_{r=0}^{\infty}
q^{-rn-r} t^{-r-1}\cr
&+\phi(q)^2
\sum_{n=0}^{\infty} (-1)^n
 q^{ n(n+1)/2}
\sum_{r=0}^{\infty}
q^{rn} t^r
\cr =&
\phi(q)^2
\sum_{n=-\infty}^{\infty} (-1)^n
\delta_n q^{ n(n+1)/2}
\sum_{r=st_n}^{\infty}
q^{rn\delta_n} t^{\delta_n r}
\cr}
\equn
$$
where $\delta_n$ and $st_n$ are defined by
$$
\eqalign{
n \geq 0  :& \delta_n =1  ,st_n=0  \cr
n <   0   :& \delta_n=-1  ,st_n=1  \cr}
\equn
$$
and where
$$
\phi(q)\equiv
\prod_{n>0}(1-q^n)
\equn
$$
Thus we can rewrite $D_{g/h}^{-1}$ as
\def\abari{\alpha}
\def\denom{
\prod_{\alpha\in S}\biggl(\sum_{n_{\alpha}=-\infty}^\infty
(-1)^{n_{\alpha}}\delta_{n_\abari}
q^{n_\abari(n_\abari+1)/2}
\sum_{r_\abari=st_{n_\abari}}^\infty
q^{r_\abari n_\abari \delta_\abari}
exp\bigl( 2 \pi i \delta_{n_\abari}r_\abari\abari\cdot z_i\bigr)\biggr)}
$$
\eqalign{
&D_{g/h}^{-1}(q,\{z_i\})=\cr
&\phi(q)^{|h|-|g|}
\denom}
\equn
$$
and obtain
the branching function,
$$
\eqalign{
b_\La^\la(q)=
&exp(2\pi i \la\cdot z)\phi(q)^{|h|-|g|}
\sum _{w \in W}\epsilon(w)e^{2\pi i\rho\cdot z}        \cr
&\sum_{\gamma \in M} q^{{1\over2}(k+g)\gamma^2+\gamma\cdot w(\La+\rho)}
exp(-2\pi iz\cdot\bigl((k+g)\gamma+w(\La+\rho)\bigr)\cr
&\denom\biggr|_{{\rm coeff\ of\ }\zeta_i^0}   }
\equn
$$
The relevant terms which contribute to
$b_\La^\la$ are those which satisfy:
$$
\la+\rho-(k+g)\gamma-\omega(\La+\rho)+\sum_{\alpha\in S}
\alpha\delta_\alpha r_\alpha=0
\; .
\equn
$$
To solve these constraints choose a basis of the lattice
spanned by the elements of $S$, say $S_0\subset S$. Let the dual be
$S^*=\{e_\alpha\}$, such that
\hbox{$e_\alpha\cdot\beta=\delta_{\alpha,\beta},$}
\ $\forall \alpha,\beta \in S_0$.
If we now define
$$
V_\beta\equiv e_\beta\cdot\biggl\{
(k+g)\gamma+\omega(\La+\rho)-\rho-\la
-\sum_{\alpha\in S\backslash S_0}\alpha r_\alpha\delta_\alpha
\biggr\}
\equn
$$
then the constraints (3.15) have solution
$$
r_\alpha =  \delta_\alpha V_\alpha \qquad \forall \alpha\in S_0
\equn
$$
which allows elimination of rank$(g)$ of the summations over the $r_\alpha$.
Since $r_\alpha\ge 0$ we have
\def\sign{\rm sign}
$\sign(V_\alpha)=\sign(\delta_\alpha)=\sign(n_\alpha), \ \
\forall \alpha\in S_0$ .

The final form of the branching function is then
$$
b_\La^\la=\phi(q)^{|h|-|g|}\sum_{w\in W}\epsilon(w)F_{w(\La+\rho)}(\tau)
\put\EQOCTone\equn
$$
where
$$\eqalign{
F_\La(\tau)&=
\sum_{\gamma\in M}
\sum_{{n_\alpha=-\infty\atop r_\alpha=st_\alpha}\atop
\alpha\in S\backslash S_0}
\sum_{{\{V_\alpha\ge0,n_\alpha\ge0\}\cup\atop
\{V_\alpha<0,n_\alpha<0\}}\atop
\alpha\in S_0}
\sigma
q^N
}
\equn
$$
and
$$
\eqalign{
\sigma&=\biggl(\prod_{\alpha\in S}\delta_{n_\alpha}\biggr)
(-1)^{\sum_{\alpha\in S}n_\alpha}\cr
N&={1 \over 2}(k+g)\gamma^2+\gamma\cdot\La
+{1 \over 2}\sum_{\alpha\in S}n_\alpha(n_\alpha+1)
+\sum_{\alpha\in S\backslash S_0}n_\alpha r_\alpha\delta_{n_\alpha}
+\sum_{\alpha\in S_0}V_\alpha n_\alpha
}
\equn
$$
and
$$
V_\alpha\equiv e_\alpha\cdot \bigl\{
(k+g)\gamma +\La-\la-\sum_{\alpha\in S\backslash S_0}
\alpha r_\alpha \delta_\alpha \bigr\}
\; .
\equn
$$
A cursory examination of this formula shows that it involves
summation over a lattice of dimension $|g|-|h|$.
This formula provides an explicit form for the branching funtions and hence
for the characters of the theory. It is however not always the most efficient
possible and in special cases,
such as $\asu(n)_k\times\asu(n)_1/\asu(n)_{k+1}$,
formulae be obtained which are considerably more efficient.
In the following sections we shall apply (3.17), evaluated by computer,
to examine the properties of the character. For large cosets, the
computer time taken to evaluate characters can become considerable.


\vskip 0.5 truecm
\secta
\noindent
{\bf \the\secno . Integer Shifts }

In coset models, as in any conformal field theory, the values of the conformal
weights of the primary fields is essential information.
For a primary field of the coset, $\phi^\La_\la$, the $h$-value
is related to that of $\La$ and $\la$ by
$$
h_\la^\La=h_\La-h_\la+n
\equn
$$
where $n$ is a non-negative integer.
The value of the integer shift $n$ is not in general known, although
its value is important. (As for example, in the construction of
four dimensional string theories based upon CFTs [\cgrovel].)
In principle, $n$ can be extracted from the character
formulae of the previous section and,
as we noted in section two, $n$ is simply the degree of the first term
in the branching function: $b^\La_\la=q^n(n_0+n_1q+\dots)$.
For illustration, consider the coset
$$
{\widehat{su}(2)_{k_1}\times \widehat{su}(2)_{k_2}}\over
\widehat{su}(2)_{k_1+k_2}
\; .
\equn
$$
For this coset the primary fields are $\phi^{\la_1\la_2}_{\la_3}$
where $\lambda_i =0,1,\cdots,k_i$.
The identification current is $\phi^{k_1,k_2}_{k_3}$, and the non-zero fields
are those
where $\la_1+\la_2-\la_3$ is even, and we have the equivalence
$\phi^{\la_1,\la_2}_{\la_3}\equiv\phi_{k_3-\la_3}^{k_1-\la_1,k_2-\la_2}$.
We determine
$n$ as a function of the parameters $\la_1,\la_2,\la_3,k_1,k_2$.
By evaluating the branching functions for a large and
representative range of these parameters, we determine
a formula for $n$
by inspection.
The form of $n$ is sufficiently simple that
this procedure is amenable, and we obtain,
for $k_1\ge k_2$,
$$
n=k_2x^2+(2s+\la_2)x+s+a+b
\equn
$$
where
$$
\def\divi{{\rm \ div\ }}
\eqalign{
s&=\bigl({{\la_3-\la_1-\la_2}\over 2} \bigr ) \; \;  \mod k_2      \cr
x&=\bigl({{\la_3-\la_1-\la_2}\over 2} \bigr ) \divi k_2      \cr
a&=(s+\la_2-k_2)\theta(s+\la_2-k_2) \cr
b&=\bigl({{\la_2-\la_1 -\la_3}\over 2}\bigr ) \theta(\la_2-\la_1-\la_3)\cr
        }
\equn
$$
For the case $k_1=k_2$ the formula is valid provided we choose
$\la_1\le\la_2$.
It is worthwhile checking that if we have :
$$
|\la_2-\la_1|\le\la_3\le\la_1+\la_2
\equn
$$
then the integer shift vanishes. The case $\la_3=\la_2+\la_1$ is immediate.
Consider the case $\la_1>\la_2$. Then setting $\la_3=\la_1+\la_2+2\alpha$,
where
$0\le\alpha<\la_2\le k_2$ it follows $s=k_2+\alpha-\la_2$,
$x=-1$, and hence $a=\alpha$, and $b=0$. Then (4.3) yields zero. The case
$\la_2\ge\la_1$ is equally trivial.
We can similarly demonstrate the invariance of
$$
h^{\la_1\la_2}_{\la_3}=n+h_{\la_1}+h_{\la_2}-h_{\la_3}
\equn
$$
under the mapping $\la_i \rightarrow k_i-\la_i\/$.
This particular coset has been studied previouly by Date et al, [\cdate]
and we reproduce their result.
In the following sections we will apply this technique to obtain
formulae for the integer shift in new examples.


 \vskip 0.5 truecm
\secta
\noindent
{\bf \the\secno . Special Case 1: $(\hat g_2)_k/\widehat{su}(3)_k$ }

In this section, the techniques of chapter three are applied  to the
particular case of $\hat g_2/\widehat{su}_3$.
This coset is fairly simple in structure, as $su(3)$ is the algebra
associated with the long root lattice of $g_2$.
We use the usual notation for $g_2$: the simple roots are
$\alpha_1$ and $\alpha_2$ where $\alpha_1^2=2$ and $\alpha_2^2={2\over3}$.
Then the $su(3)$ has positive roots $\alpha_1,\alpha_1+3\alpha_2$
and $2\alpha_1+3\alpha_2$. For this coset, unusually, there are
no identification currents. If we label the fields $\phi_\la^\La$,
where $\La$ and $\la$ label irreducible representations
of $g_2$ and $su(3)$ respectively,
then all pairs $(\La,\la)$ are allowed and inequivalent.
In evaluating the
branching function formula, \EQOCTone , we find (3.15) becomes
$$
\lambda+\sum_{\alpha\in S}\alpha\delta_{n_{\alpha}}r_{\alpha}
+\rho-(k+g)\gamma-w(\La+\rho)=0
\equn
$$
The summation over $S$ is simply
the summation over $g_2^+\backslash su(3)^+$,  the short
positive roots of $g_2$, which is $\{\alpha_2,\alpha_{12},\alpha_{12^2}\}$,
where $\alpha_{12^2}\equiv\alpha_1+2\alpha_2$.
We use these constraints to eliminate two of the summations over $r_\alpha$
from (3.19). Letting $r_2\equiv r_{\alpha_2}$ etc.,we find
$$
\eqalign{
\delta_2r_2&=-\delta_{12^2}r_{12^2}-\alpha_{12^2}\cdot V
\cr
\delta_{12}r_{12}&=-\delta_{12^2}r_{12^2}-\alpha_{1^22^3}\cdot V
}
\equn
$$
where
$$
V\equiv\lambda+\rho-w(\Lambda+\rho)-(k+g)\gamma
\equn
$$
Thus, given $n_{12^2},r_{12^2},w,\gamma$, we can fix $r_2$,$r_{12}$ and the
sign of $n_2$ and $n_{12}$.
The explicit formula for the branching functions is then,
$$
b^\lambda_\La(q)=
\phi(q)^{-6}
\sum_{w\in{W_{g_2}}}
\sum_{\gamma\in M}
\sum_{n_{12^2}\in Z}
\sum_{r_{12^2}=st_{12^2}}^\infty
\sum_{ \{V_2\geq 0,n_2\geq0\}\atop
      \cup\{V_2<0,n_2<0\}}
\sum_{ \{V_{12}\geq 0,n_{12}\geq0\}\atop
      \cup\{V_{12}<0,n_{12}<0\}}
\sigma q^N
\equn
$$
where
$$
\eqalign{
\sigma&=\epsilon(w).(-1)^{n_2+n_{12}+n_{12^2}}
\delta_{n_2}\delta_{n_{12}}\delta_{n_{12^2}}\cr
N&= r_2n_2\delta_2+r_{12}n_{12}\delta _{12}
   + r_{12^2}n_{12^2}\delta_{12^2}
   +{1\over 2}(k+g)\gamma^2+\gamma\cdot w(\Lambda+\rho)
\cr&{}+n_2(n_2+1)/2+n_{12}(n_{12}+1)/2+n_{12^2}(n_{12^2}+1)/2
}
\equn
$$

Consider the theory at level $1$. The central charge is
$c=4/5$ and the non-zero weight labels are $(0,1)$ and $(1,0)$
for $su_3$, both at spin $1/3$, and $(0,1)$ for $g_2$, at spin $2/5$.
We therefore have six fields and no identification currents.
The weights of the fields are $h_\La^\mu=h_\La-h_\mu+n$,where $n$
is the degree of the first term in the corresponding
branching function.
The computed branching functions are:
$$
\eqalign{
b_{(0,0)}^{(0,0)}&=1+q^2+2q^3+3q^4+4q^5+7q^6+\ldots
\cr
b_{(0,0)}^{(0,1)}&=q+q^2+2q^3+2q^4+4q^5+5q^6+\ldots
\cr
&=b_{(0,0)}^{(1,0)}
\cr
b_{(0,0)}^{(0,1)}&=1+2q+2q^2+4q^3+5q^4+8q^6 +\ldots
\cr
b_{(0,1)}^{(0,1)}&=1+q+2q^2+3q^4+5q^5+7q^6+\ldots
\cr
&=b_{(0,1)}^{(1,0)}
\cr}
\equn
$$
Hence we obtain fields with weights:
$h=0,{2 \over 5},({2 \over 3})^2,({1 \over 15})^2\/$. The weights correspond
to the NDMI of the $c={4 \over 5}$ minimal model [\ccappelli].
This well known example, corresponds to the first element of the $W_3$
algebras.
Thus we have the equivalence
$$
{ {\hat g}{}_2 )_1 \over {\asu(3) }_1  }
\equiv
{ {\asu(3) }_1 \times {\asu(3) }_1 \over {\asu(3) }_2 }
\equiv
{ {\asu(2) }_2 \times {\asu(2) }_1 \over {\asu(2) }_4 }
\Bigl|_{NDMI}
\; .
\equn
$$
Examining the identity character, one can observe an additional state at
level three corresponding
to the spin-3 field which generates the $W_3$ symmetry.
( For a pure Virasoro algebra the representation with $h=0$ generally has
only a single state
at level 3, $L_{-3}| 0>$, since the state $L_{-1} |0>$ is null.)
For a coset theory,
the algebra is generally an extension of the Virasoro algebra [\cextended].
Examining the zero character is an easy way to discover the spin of the lowest
extra field in the symmetry algebra.

We present a formula for the integer shifts
using the same technique as for the diagonal
$\asu(2)$ case.
As in that case we only have five
parameters, and hence the solution is not hard to obtain. We find
for a primary field $\phi^{\mu}_{\La}$ the integer shift is
\def\ka{\kappa}
$$
\eqalign{
n=&\ka_1\theta(\ka_1)\theta(\ka_2) \; +\ka_2\theta(\ka_2)
\; +(\la_1-\mu_1-\mu_2)\theta(\la_1-\mu_1-\mu_2)             \cr
&+(\ka_1+\ka_2)\theta(\ka_1+\ka_2)\theta(\la_1+\la_2-1-\mu_2)}
\equn
$$
where
$$
\Lambda=\la_i w_i\quad
\mu=\mu_i\nu_1\quad
\mu-\Lambda=\ka_i\nu_i\quad
\mu_1\le\mu_2
\equn
$$
having chosen $w_i$ and $\nu_i$ as the standard fundamental weights of
$g_2$ and $su(3)$ respectively. (The $\theta$-function is the usual
step function with $\theta(x)=0$ for $x<0$ and $\theta(x)=1$ for
$x\geq 0$.)
Note that when $\mu-\Lambda$ has positive Dynkin components in the $su(3)$
basis then the formula reduces to $(\mu-\Lambda)\cdot\psi$, where $\psi$
is the highest root.


\vskip .4 truecm
\secta
\noindent
{\bf \the\secno . Special Case 2: $\widehat{su}(3)_k/\widehat{su}(2)_{4k}$ }

This very interesting example of a coset CFT provides
a counter example to some of the preconceptions relating to coset spaces.
For $k=1$ the coset is the trivial theory. This is related to the
conformal embedding of $\widehat{su}(2)_4$ within $\asu(3)_1$
[\cravanini,\cextended].
For $k>2$ the coset is a perfectly normal example of a theory
where the non-zero characters and equivalence between characters
is determined by an identification current.  This current is
given by
$$
J_I = \phi^{(0,0)}_{4k}
\equn
$$
and the action of $J_I$ upon a field is
$$
J_I \cdot \phi^{\La}_l = \phi^{\La}_{4k-l}
\equn
$$
Using this, the fields $\phi^{\La}_{l}$ are non-zero provided
$l$ is even and we have the equivalence $\phi^{\La}_l\equiv\phi^{\La}_{4k-l}$.
The non-zero condition is completely equivalent to demanding that
$P\La-l$ lies within the root lattice of $su(2)$,
where $P\La$ is the $su(2)$ weight $\La$ is mapped to
by the embedding.
For $k=3$, $c=10/7$ and it can be shown that this coset is equivalent to the
third member of the $W_3$ minimal series. This provides a check that
the identification current does entirely specify the spectrum for $k>2$.

For $k=2$ however we find we have a \lq\lq maverick\quotesr\ coset theory.
For this coset $c=4/5$. Hence this coset must correspond to the
minimal model with this $c$-value. There are two possibilities
$-$either it corresponds to the  diagonal modular invariant or it
corresponds to the non-diagonal modular invariant [\ccappelli]
(which is equivalent
to the first element of the $W_3$ minimal series). However, if we consider
the potential field $\phi^{(0,0)}_2$ say, then this has
$h=4/5$ which does not lie within the spectrum of the $c=4/5$ minimal model.
To investigate this example further we must give a formula for the
characters of $\widehat{su}(3)_k/\widehat{su}(2)_{4k}$.
With the standard conventions
the embedding of $\widehat{su}(2)_{4k}$ within $\widehat{su}(3)_k$ is given by
$$
J^{+}(z)= \sqrt{2}(  J^{1+2}(z)+ J^{-2}(z) ) ; \hskip 0.5 truecm
J^{-}(z)= \sqrt{2}(  J^{-1-2}(z)+ J^{2}(z)  ) ; \hskip 0.5 true cm
J^{0}(z)=  2 H^1(z)
\equn
$$
where the indices refer to the simple roots $\alpha_1$ and $\alpha_2$.
Using the techniques of section 3, we can deduce a formula for a
character $\chi^{\La}_{l}$,
$$
\chi^{\La}_l=
\phi(\tau)^{-5} \sum_{w\in W_{SU(3)}}
\sum_{\gamma\in M}
\sum_{n_1=-\infty}^{\infty}\sum_{r_1=st(n_1)}^{\infty}
\sum_{{\{V<0,n_2<0\}} \atop\{ V\geq 0,n_2\geq 0\}  }
(-1)^{n_1+n_2-1} \delta_{n_1}\delta_{n_2}
q^{N(\gamma,n_i,r_1,V)}
\equn
$$
where
$$
\eqalign{
V=& ( w(\La+\rho)-\rho )\cdot \alpha_1 -{ l \over 2}
-2r_1\delta_1 -1  +(k+g) \gamma \cdot \alpha_1  \cr
N=& {1 \over  2} (k+g) \gamma^2 +\gamma\cdot w(\La+\rho)
+{1\over 2} n_1(n_1+1)
+\delta_1 n_1 r_1
+{1\over 2} n_2(n_2+1)
+\delta_2 n_2 V
\cr}
\equn
$$
If we evaluate this character for $k=2$ then we find more characters
are zero than we would expect from the identification current
and there exist more equivalences than one would expect.
The equivalences of the characters are
given in table~1. From this, and comparing with the minimal model,
we find there are more zero characters, and also there are more equivalences
than expected. Roughly, speaking there being one extra equivalence so
that $\chi^{0}_{00}$ is eqivalent to $\chi_{11}^{4}$ in addition to the
expected $\chi^{8}_{00}$.
Taking these into account we find the spectrum of $h$-values matches
that of the NDMI of the $c=4/5$ minimal model ( $h=0,2/5,(2/3)^2,(1/15)^2$).
This is also the first element of the $W_3$ minimal series so this coset
has an extended algebra containing a spin-3 field as in the
 $\hat{g}_2/\widehat{su}(3)$
case considered earlier.
This maverick coset space is related to the NDMI which exists for
$\widehat{su}(3)_2\times \widehat{su}(2)_8^*$,
$$ \eqalign{
Z= &\Bigl| \chi^{(00)}_0 +\chi^{(00)}_8 +\chi^{(11)}_4 \Bigr|^2
+\Bigl| \chi^{(11)}_2 +\chi^{(11)}_6 +\chi^{(00)}_4 \Bigr|^2
+\Bigl| \chi^{(10)}_2 +\chi^{(10)}_6 +\chi^{(02)}_4 \Bigr|^2  \cr
&+\Bigl| \chi^{(01)}_2 +\chi^{(01)}_6 +\chi^{(20)}_4 \Bigr|^2
+\Bigl| \chi^{(20)}_0 +\chi^{(20)}_8 +\chi^{(01)}_4 \Bigr|^2
+\Bigl| \chi^{(02)}_0 +\chi^{(02)}_8 +\chi^{(10)}_4 \Bigr|^2
\cr} \equn
$$
As we can see this NDMI is generating the characters of the coset theory
in an analogous manner to that in which the
identification current works normally. However this case is
not generated by any simple current,
as can be seen, for example, by looking at the component
$\chi^{00}_0+\chi^{00}_8+\chi^{11}_4$. The last term is not related to
the others by any simple current.

We feel this single example to be very instructive. Although the
technology of Schellekens and Yankielowicz provides a means of
determining the primary fields of a coset theory in most cases
this is an example where it does not work.  This method relies on several
assumtions, clearly stated as e.g. in ref.[\cshellb], which may
not hold. In particular,
this theory is a counterexample to the proposition
that characters will be non-zero provided that
$P\La-\la$ lies within
 the lattice $P\calm_g$,($P$ denoting the embedding of the weight space),
 as can be seen by examining the
character $\chi^{00}_2$ which is zero for $k=2$.
The equivalence
of characters is also clearly greater in this case than that expected
by considering the automorphisms of the extended dynkin diagrams of $\hat g$
and $\hat h$.
Maverick coset spaces will be dealt with in
a future work [\cfuture].
Coset theories provide a rich framework in which to
construct rational CFTs which is not yet fully understood.


\baselineskip = 15pt

\vskip .4 truecm
\secta
\noindent
{\bf \the\secno . Conclusions  }
Coset conformal field theories have proved a valuable tool in
the study of rational conformal field theories.
In this paper we have studied the characters
of a coset conformal field theory and produced
some quite general formulae for these characters.
These characters determine the spectrum of
the coset theories and as such are crucial in determining
the physical spectrum in string theories.
In particular the spectrum of the $h-$values is determined by the characters,
including the integer shifts. We obtained a formula for the integer shift
agreeing with calculations done previously for specific cases.
The characters can also be used to study the existence of
extended symmetries in coset CFTs.
We have also used
these character formulae to provide an example of a coset theory
which cannot be described by the identification current method.
This coset $\asu(3)_2/\asu(2)_8$, is a \lq\lq maverick'' theory which has
more zero characters and equivalences than specified by the identification
current method.
We expect the study of this and other maverick theories,
via the branching function formulae or other methods to
provide much insight into the study of coset CFTs.

This work was supported by an S.E.R.C. advanced fellowship (D.C.D.)
and an S.E.R.C studentship (K.G.J.).  One of us
(D.C.D.) wishes to thank
Zvi Bern for useful discussions and a N.A.T.O collaborative
grant CRG-910285.


\Appendix {Generalising the Character formulae}
Suppose we have a coset $\hat g / \hat h$ which decomposes into the
form:
$$
{\hat g_k \over \hat h_l}={{(\hat g_1)_{k_1}\times \dots \times(\hat
g_n)_{k_n}}
\over
{(\hat h_1)_{l_1}\times \dots \times (\hat h_m)_{l_m}}}
\equn
$$
where $\{ G_i \} $ and $\{H_{\jbar} \} $ are semisimple compact Lie groups
other than $U(1)$.
Note that this means that for each  $\jbar$, $H_\jbar \subset G_{i(\jbar)}$,
for some $i(\jbar)$, which will be an aid to the calculation in what follows.
In this appendix we are using barred indices to refer to the subalgebras $\hat
h$
and unbarred indices for the algebras $\hat g$.
The embedding of $h$ in $g$ is specified by
$$
\eqalign{
(H^{h_{\ibar}})^\jbar&=P^{\ibar\jbar}_{kl}(H^{g_k})^l\cr
E^{\alpha_\ibar}&=M^{\alpha_\ibar}_{\alpha_k}E^{\alpha_k}\cr
}
\equn
$$
where $(H^{g_k})^l$ denotes an element of the CSA of $g_k$ and
$\alpha_k$ denotes a root of $g_k$.
Associated with each Cartan generator $(H^{g_k})^l$ there
is a basis vector of the root space $e_l^k$.
Denoting by $\cal M$ the root lattice and $\cal W$ the weight lattice,
let elements of ${\calm}_g$ be
$\alpha\equiv(\alpha_1,\dots,\alpha_n)$
we have the mapping:
\def\calm{{\cal M}}
$$\eqalign{
P:\calm_g&\rightarrow{\cal W}_h\cr
\alpha&\rightarrow\bar\la=
P^{\ibar\jbar}_{kl}(\alpha_k\cdot e_l^k)e_\jbar^\ibar\cr
}
\equn
$$
An arbritary vector in the weight space of $\hat h$ may be written
$$
v \equiv -2\pi i \sum_{\ibar=1}^m
\bigl(\sum_{\jbar=1}^{{\rm rank }(h_\ibar)}z^\ibar_\jbar e^\jbar_\ibar
+\tau(\Lambda_0)_\ibar +u_\ibar\delta_\ibar\bigr)
\equn
$$
Again writing $\L=(\La_1,\dots,\La_n)$,
where $\La_i$ is a highest weight label for $\hat g_i$,
the branching function equation becomes
$$
ch_\La^g(v)=\sum_\mu b_\La^\mu ch_\mu(v)
\equn
$$
where $\mu$ are weight labels for $\hat h$.
In evaluating the lhs we can use (3.1),(3.2),(3.3) directly,
provided we set
$$
z_k^l=P^{\ibar\jbar}_{kl}z^\ibar_\jbar
\equn
$$
When evaluating $D_g$ using (3.2) we can use the following relation
$$
\alpha_k\cdot z_k=\alpha_\ibar\cdot z_\ibar\qquad\qquad\forall\
\alpha_k,\alpha_\ibar\  {\rm st.}\ \ M^{\alpha_\ibar}_{\alpha_k}\neq 0
\equn
$$
It is therefore easy to decompose $D_g$:
$$
D_{\hat g_{j(\ibar)}}=D_{\hat h _{\ibar_1}}\dots D_{\hat h_{\ibar_p}}
\prod_{\alpha \in S}(1-e^{-\alpha})^{{\rm mult} \alpha}
\equn
$$
where $H_{\ibar_1}\times \dots\times H_{\ibar_p} \subset G_j$, and $S$
is the complement of the set of positive roots of the $ h_{\ibar} $ in
those of $Pg_{j_(\ibar)}$. It is important to note (A.8) is meaningful only
when evaluated on the weight space of $h$ using (A.6). Having thus removed
$D_{\hat h}$ and
writing $D_{\hat g}=D_{\hat g/\hat h}D_{\hat h}$ we have
$$
{\prod_{i=1}^n N_{\Lambda_i}^{\hat g_i}(q,z_i)
\over
D_{\hat g / \hat h}(q,\{z_\jbar\})}
=\sum_\mu b_{\Lambda_1\dots\Lambda_n}^{\mu_1\dots\mu_m}
\prod_{\jbar=1}^m N_{\mu_\jbar}^{\hat h_\jbar}(q,z_\jbar)
\equn
$$
It is convenient to define the variables
$$
\hfill y^\ibar_\jbar\equiv exp(2 \pi iz^\ibar_\jbar)
\qquad     \hfill
\ibar=1\dots m,\quad\jbar=1\dots{\rm rank(}h_\ibar{\rm )}
\equn
$$
Thus, as in section 3, we obtain
\def\bfl{b_{\La_1\dots\La_n}^{\la_1\dots\la_m}(q)}
$$
\bfl
={exp\biggl(2\pi i(\sum_{\ibar=1}^m\lambda_\ibar\cdot z_\ibar)\biggr)
{\prod_{i=1}^nN_{\Lambda_i}^{\hat g_i}(q,z_i)
\over
D_{\hat g / \hat h} (q,\{z_\ibar\})}}\biggl\vert_{{\rm coeff}\atop
{\rm of\ } (y^\ibar_\jbar)^0}
\equn
$$
Again, $\delta_n$ and $st_n$ are defined by
$$
\eqalign{
n \geq 0  :& \delta_n =1  ,st_n=0  \cr
n <   0   :& \delta_n=-1  ,st_n=1  \cr}
\equn
$$
\def\denom{
\prod_{\bar\alpha_i\in S}\biggl(
\sum_{n_{\bar\alpha_i}=-\infty\atop r_\abari=st_{n_\abari}}^\infty
(-1)^{n_{\bar\alpha_i}}\delta_{n_\abari}
q^{n_\abari(n_\abari+1)/2}
q^{r_\abari n_\abari \delta_\abari}
exp\bigl( 2 \pi i \delta_{n_\abari}r_\abari\abari\cdot z_i\bigr)\biggr)}
Thus (3.2),(3.4),(3.13),(A.11)  together give us our general result for the
branching
function
$$
\eqalign{
\bfl = &exp(2\pi i(\sum_\bari\lambda_\bari\cdot z_\bari))\phi(q)^{|h|-|g|}
\prod_{i=1}^n\biggl\{\sum_{w_i\in W_{g_i}}\epsilon(w_i)
e^{2\pi i\rho_i\cdot z_i}
\cr
&\times\sum_{\gamma_i\in M_i^L}
q^{{1\over2}(k_i+g_i)\gamma_i^2+\gamma_i\cdot w_i(\Lambda_i+\rho_i)}
exp\bigl(-2\pi i(k_i\gamma_i\cdot z_i+w_i(\Lambda_i+\rho_i)\cdot z_i)\bigr)
\biggr\}
\cr
\times&\denom\Bigl\vert_{{\rm coeff}\atop {\rm of\ }(y_\ibar)_\jbar^0}}
\equn
$$
The relevant terms satisfy
$$
\la+P\sum_{\alpha\in S}\alpha \delta_{n_\alpha}r_\alpha+P\rho=
P\bigl\{\sum_{i=1}^n(k_i+g_i)\gamma_i+w_i(\La_i+\rho_i)\bigr\}
\equn
$$
Note this this equation implies the well known condition
$\la-P\La\in P\calm_g$.
Thus using (A.13) directly we may compute any branching function.
In order to make such a computation efficient it is necessary to
solve the constraints (A.14) for some of the $r_{\bar\alpha_k}$
so as to eliminate them from (A.13).
To do this define $S_0\subset PS$ form a basis for $P\calm_g$, and
let $S_0^*=\{w_\alpha\}$ be its dual, such that
$w_\alpha\cdot P\beta=\delta_{\alpha,\beta}$.
Define
$$
\eqalign{
V&=P\biggl(\sum_k\bigl(k_k+g_k)\gamma_k  +w_k(\La_k+\rho_k)
-\rho_k\bigr)
-\sum_{\alpha\in S\backslash S_0}\alpha\delta_{n_\alpha}r_\alpha\biggr)-\la\cr
V_\beta&\equiv w_\beta\cdot V\cr
}
\equn
$$
Thus the constraints (A.14) become
$$
V_\beta =r_\beta\delta_{n_\beta}
\equn
$$
We can then  insert this into (A.13) to  obtain the general result:
$$
\bfl=\phi(q)^{|g|-|h|}
\prod_{i=1}^n
\biggl\{\sum_{w_i\in W_{g_i}}
\epsilon(w_i)F_{w,\L}(q)
\equn
$$
where
$$
F_{w,\La}(q)=\sum_{\gamma_i\in M_i}
\sum_{{n_\alpha =-\infty \atop r_\alpha =st_\alpha}
\atop \alpha \in PS\backslash S_0}
\sum_{{\{V_\alpha\ge0,n_\alpha\ge0\}\cup\atop
\{V_\alpha<0,n_\alpha<0\}}\atop
\alpha\in S_0}
\sigma q^N
\equn
$$
where
$$
\eqalign{
\sigma&=\biggl(\prod_{\alpha\in S}\delta_{n_\alpha}\biggr)
(-1)^{\sum_{\alpha\in S}n_\alpha}\cr
N&={1\over2}\sum_i (k_i+g_i)\gamma_i^2+w_i(\La_i+\rho_i)\cdot\gamma_i\cr
&+{1\over2}\sum_{\alpha\in S}n_\alpha(n_\alpha+1)
+\sum_{\alpha\in PS\backslash S_0}n_\alpha r_\alpha \delta_{n_\alpha}
+\sum_{\alpha\in S_0}V_\alpha n_\alpha\cr
}
\equn
$$

\vfill\break

$$
\vbox{\offinterlineskip
\hrule
\halign{ &\vrule# & \strut\quad\hfil#\quad\cr
height2pt&\omit&&\omit&&\omit&&\omit&\cr
& Equivalences of Character   &&Character Value\qquad\qquad&\cr
height2pt&\omit&&\omit&&\omit&&\omit&\cr
\noalign{\hrule}
height2pt&\omit&&\omit&&\omit&&\omit&\cr
&\quad$\chi_{0}^{(00)} \equiv \chi_{8}^{(00)}\equiv \chi_{4}^{(11)}\hfill$
&& $1 ~~~~~~ +q^2 +2q^3 +3q^4 +4q^5 \cdots  $   &\cr
&\quad  $\chi_{2}^{(11)} \equiv \chi_{6}^{(11)}\equiv \chi_{4}^{(00)}\hfill$
&& $ 1+2q +2q^2 +4q^3 +5q^4 +8q^5 \cdots $   &\cr
&\quad $\chi_{2}^{(10)} \equiv \chi_{6}^{(10)}\equiv \chi_{4}^{(02)}\hfill$
&& $1 +q +2q^2 +3q^3+5q^4 +7q^5 \cdots $   &\cr
&\quad $\chi_{2}^{(01)} \equiv \chi_{6}^{(01)}\equiv \chi_{4}^{(20)}\hfill$
&& $1 +q +2q^2 +3q^3+5q^4 +7q^5 \cdots $    &\cr
&\quad $\chi_{0}^{(20)} \equiv \chi_{8}^{(20)}\equiv \chi_{4}^{(01)}\hfill$
&&
$1 +q +2q^2 +2q^3 +4q^4 +5q^5 \cdots $
&\cr
&
$\quad \chi_{0}^{(02)} \equiv \chi_{8}^{(02)}\equiv \chi_{4}^{(10)}\hfill$
&&
$1 +q +2q^2 +2q^3 +4q^4 +5q^5 \cdots $
&\cr
&\quad $\chi_{2}^{(00)} \equiv \chi_{6}^{(00)}\hfill $
&&$~$ 0 \hfill  &\cr
&\quad  $\chi_{2}^{(20)} \equiv \chi_{6}^{(20)}\hfill $
&&$~$ 0 \hfill  &\cr
&\quad  $\chi_{2}^{(02)} \equiv \chi_{6}^{(02)}\hfill $
&&$~$ 0 \hfill  &\cr
&\quad  $\chi_{0}^{(10)} \equiv \chi_{8}^{(10)}\hfill $
&&$~$ 0 \hfill  &\cr
&\quad $\chi_{0}^{(01)} \equiv \chi_{8}^{(01)}\hfill $
&&$~$ 0 \hfill  &\cr
&\quad  $\chi_{0}^{(11)} \equiv \chi_{8}^{(11)} \hfill$
&&$~$ 0 \hfill  &\cr
height2pt&\omit&&\omit&&\omit&&\omit&\cr}
\hrule}
$$
{\bf Table 1.} For the coset theory
${\widehat{su}(3) }_2/{\widehat{su}(2)}_8$,
we show the extra equivalences and vanishing of characters beyond
that expected by the identification current method.
Using the identification currrent method all characters shown
are expected
to be non-zero.

\vfill\break
\noindent
{\bf References }

\parindent=-13 pt
\parskip=0 pt

[\cgko] P.\ Goddard, A.\ Kent, D.\ Olive., Phys. Lett.
 {\bf 152} (1985) 88;
 Commun. Math. Phys. {\bf 103} (1986) 105.

[\cbpz]  A.A.\ Belavin, A.M.\ Polyakov and A.B.
Zamolodchikov,  Nucl.\ Phys.\ {\bf B241}(1984) 333;
J.\ L.\ Cardy, Nucl. Phys. {\bf B240} (1984) 514.

[\cfms]
D.\ Friedan, Z.\ Qiu and S.\ Shenker,
Phys. Rev. Lett. {\bf 52} (1984) 1575.

[\ccardya]
J. Cardy, Nucl.\ Phys.\ {\bf B270}  (1986) 186.

[\cgepwitt]
D.\ Gepner and E. Witten,
Nucl.\ Phys.\ {\bf B278} (1986) 493.

[\ckac]
V.G.\ Kac, Func.\ Anal.\ App.
{\bf 1} (1967) 328;
R.V.\ Moody, Bull.\ Am.\ Math.\ Soc.\
{\bf 73} (1967) 217;
K.\ Bardakci and M.B.\ Halpern, Phys.\ Rev.\
{\bf D3} (1971) 2493.

[\cgod] P. Goddard and D. Olive, Int. J. Mod. Phys. {\bf A1} (1986) 303.

[\cgrovel]
D.\ Gepner Nucl. Phys. {\bf B296} (1988) 757;
A.Font, L.\ Ibanez and F.\ Quevedo , Phys. Lett. {\bf 217B} (1989) 272;
D.\ Bailin, D.C.\ Dunbar and A.\ Love Int. J. Mod. Phys.
 {\bf A6} (1991) 1659.

[\cmoore]
G. Moore and N. Sieberg, Phys. \ Lett.\ {\bf 220B} (1989) 422;
D.\ Gepner, Phys.\ Lett.\ {\bf 226B} (1989) 207;
W.\ Lerche, C.\ Vafa and N.\ Warner, Nucl.\ Phys.\ {\bf B324} (1989) 673.

[\cshellb]
A.N.\ Schellekens and S.\ Yankielowicz,
Nucl. Phys. {\bf B334} (1990) 67.

[\czam] A.B. Zamolodchikov, Theor. Math. Phys. {\bf 65} (1986) 1205.

[\cverlinde]
E.\ Verlinde, Nucl.\ Phys.\ {\bf B300} (1988) 360.

[\cshella]
A.N.\ Schellekens and S.\ Yankielowicz,
Nucl. Phys. {\bf B327} (1989) 673;
Phys.\ Lett. {\bf 227B} (1989) 387.

[\cintr]
A.K.\  Intriligator, Nucl. Phys. {\bf B332} (1990) 541.

[\cbais]
F. Bais, P. Bouwknegt, K. Schoutens and M. Surridge,
Nucl. Phys. {\bf B304} (1988) 348.

[\cdate]  E.\ Date, M.\ Jimbo, A.\ Kuniba, T.\ Miwa, M.\ Okado,
Nucl. Phys. {\bf B290} (1987) 231.

[\ckaca]
V.\ Kac and D.H.\ Petersen, Adv. Math. {\bf 53} (1984) 125;
V.\ Kac and M.\ Wakimoto, Adv. Math. {\bf 70} (1988) 156.

[\ccappelli]
A.~Cappelli, C.~Itzykson and J.-B.~Zuber, Nucl.~Phys.\ {\bf B280}
(1987) 445.

[\cravanini] F.\ Ravanini, Int. J. Mod. Phys. {\bf A3} (1988) 397;
P.\ Christe and F. Ravanini , Int. J. Mod. Phy. {\bf A4} (1989) 897.

[\cextended] Z.\ Bern and D.C.\ Dunbar, Phys. Lett. {\bf 248B} (1990) 317.

[\cfuture] D.C.\ Dunbar and K.G.\ Joshi, in preparation


\vfill\eject
\bye